\documentclass{appolb}
\usepackage{graphicx}
\usepackage{amssymb}
\usepackage{color}
\begin{document}

\title{Level spacing distribution \\for the prototype of the Bianchi IX model\thanks{Presented
at the conference \emph{Random Matrix Theory: Foundations and Applications}, Krak\'{o}w, July 1-6, 2014.}}
\author{Jakub Mielczarek$^{ab}$\thanks{jakub.mielczarek@uj.edu.pl},
W{\l}odzimierz Piechocki$^{a}$\thanks{piech@fuw.edu.pl}
\address{$^{a}$Theoretical Physics Division, National Centre for Nuclear Research, \\Ho{\.z}a 69,
00-681 Warsaw, Poland \\
$^{b}$Institute of Physics, Jagiellonian University, {\L}ojasiewicza 11, 30-348 Cracow, Poland}}

\maketitle
\begin{abstract}
Our results concern quantum chaos of the vacuum Bianchi IX model.
We apply the equilateral triangle potential well approximation to
the potential of the Bianchi IX model to solve the eigenvalue
problem for the physical Hamiltonian. Such approximation is
well satisfied in vicinity of the cosmic singularity. Level
spacing distribution of the eigenvalues is studied with and
without applying the unfolding procedure. In both cases, the
obtained distributions are qualitatively described by  Brody's
distribution with the parameter $\beta\approx0.3$, revealing some
sort of the level repulsion. The observed repulsion may reflect
chaotic nature of the classical dynamics of the Bianchi IX
universe. However, full understanding of this effects will require
examination of the Bianchi IX model with the exact potential.

\end{abstract}
 \vspace{0.5cm}

\section{Introduction}

Astronomical observations tell us that the Universe has been
expanding for almost 14 billion years so it must have emerged from
a state with extremely high energy densities of matter fields.
Theoretical attempts of understanding the early stage of
the Universe are, however, plagued by the cosmological singularity
characterized by blowing up of physical invariants. This concerns
almost all known general relativity models of the universe
(Friedmann, Lema\^{i}tre, Kasner, Bianchi, Szekeres, \dots). The
existence of the cosmological singularities in solutions to
Einstein's equations means that this classical theory is
incomplete. Expectation is that quantization may heal the
singularities.

The  Friedmann-Robertson-Walker (FRW) model, which is
homogenous and isotropic, is commonly used to describe the
cosmological data. However, the isotropy is dynamically unstable
in the backward evolution towards the cosmic singularity
\cite{LK}. The general anisotropic model including FRW as a
special case is the Bianchi IX model.  It is the best prototype
for the Belinskii, Khalatnikov and Lifshitz (BKL) scenario, which
is thought to be a generic solution to the Einstein equations
near the cosmological singularity \cite{BKL1,BKL2}.

The classical dynamics of the Bianchi IX model is both singular
\cite{BKL1,cwm2} and chaotic \cite{BKL1,cwm2,NJC}. In this
article we look for symptoms of the chaotic behavior at the
quantum level. For this purpose, the spectrum of  canonically
quantized Hamiltonian for the vacuum Bianchi IX model is analyzed.
In our considerations, the hard wall approximation (which is well
satisfied in vicinity of the cosmic singularity) for the Bianchi
IX potential is applied.

\section{The Bianchi IX model}

The Bianchi IX model can be viewed as a generalization of the
positively curved ($k=1$) FRW with two additional degrees of
freedom parameterizing the anisotropy \cite{DHK}. The line element
of the model reads
\begin{equation}\label{eq}
ds^2= - N^2(t)\; dt^2+\frac{1}{4}\sum_{i=1}^{3} q^2_i(t){\omega^i}\otimes{\omega^i}\, ,
\end{equation}
where $\omega^i$ are 1-forms satisfying $d \omega^i=\frac{1}{2}\epsilon_{ijk}\omega^j\wedge\omega^k$.
The $N(t)$ is a lapse function and the $q_i(t)$ function are usually parametrised as $q_i(t)=a(t)e^{\beta_i}$,
where $\sum _{i=1}^{3} \beta_i =0$.  The $a=\left(q_1q_2q_3\right)^{1/3}=e^{-\Omega}$ is an average
scale factor  related to $\Omega$, playing a role of internal time.  Based on the $\beta_i$ functions, it is
convenient to define parameters $\beta_{+}:=\beta_1+\beta_2$ and $\beta_{-}:=(\beta_1-\beta_2)/\sqrt{3}$.
The closed ($k=1$) FRW model is recovered for $\beta_{+}=0=\beta_{-}$, with $a(t)$ corresponding to the
FRW scale factor.

The   action integral for Bianchi type models reads
\cite{cwm2,cwm3}
\begin{equation}\label{a9}
    I = \int (p_+ d \beta_+ + p_- d \beta_- - H d \Omega ),
\end{equation}
where $\beta_\pm$ and $p_\pm$  are canonical variables,
$\left\{\beta_{\pm},p_{\pm} \right\} = \delta_{\pm,\pm}$, whereas
$H = H (\Omega, \beta_\pm, p_\pm )$ is the physical (true)
Hamiltonian. The relation of the internal time $\Omega$ to the
cosmic time $t$ reads: $NHdt =-\sqrt{2/3\pi} \exp(-3 \Omega)\; d\Omega$ \cite{cwm3}.

\begin{figure}[htb]
\centerline{
\includegraphics[width=6cm, angle = 0]{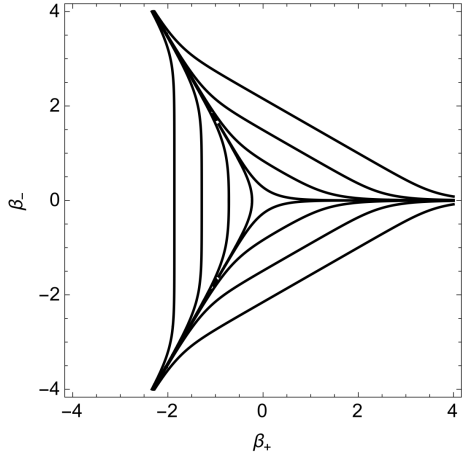}}
\caption{Equipotential lines for the Bianchi IX potential.}
\label{figure1}
\end{figure}

The Hamiltonian in (\ref{a9}) has the form
\begin{equation}\label{b4}
H^2 = p_+^2 + p_-^2 + e^{-4 \Omega}\,(V - 1),
\label{ClassicalHamiltonian}
\end{equation}
where the potential $V$ for the Bianchi IX model is defined to be
\begin{equation}\label{b5}
V =  - \frac{4}{3}e^{-2\beta_+}\cosh(2\sqrt{3}\beta_-) +
\frac{2}{3}e^{4\beta_+}(\cosh(4\sqrt{3}\beta_-) -1) +
\frac{1}{3}e^{-8\beta_+}  +1 .
\end{equation} 
As seen from Fig. \ref{figure1}, the potential is characterized by
$C_{3v}$  symmetry with equipotential lines asymptotically
arranging in the form of the equilateral triangle.

\section{Quantization}

In order to quantize the system, the canonical procedure of
quantization is applied:
\begin{equation}
\left\{\beta_{\pm},p_{\pm} \right\} = \delta_{\pm,\pm} \rightarrow
\left[ \hat{\beta}_{\pm},\hat{p}_{\pm}\right] = i \hslash \delta_{\pm,\pm} \hat{\mathbb{I}}.
\end{equation}
Using the Schor\"{o}dinger representation for the above commutation relation, the
quantum  operator corresponding to (\ref{ClassicalHamiltonian}) reads:
\begin{equation}\label{ap6}
\widehat{H^2}= -\hslash^2\Big(\frac{\partial^2}{\partial \beta^2_+} +
\frac{\partial^2}{\partial \beta^2_-}\Big) + e^{-4 \Omega } \left(V(\beta_\pm) - 1\right).
\end{equation}
The domain $D(\widehat{H^2})$ is chosen in such a way that the
operator $\widehat{H^2}$ is self-adjoint, where $D(\widehat{H^2})
\subset \mathcal{H} := L^2 (S_{\Omega}, d\mu)$ with the subset
$S_{\Omega}$ defined as:
\begin{equation}\label{ap52}
S_{\Omega} := \left\{(\beta_-, \beta_+) \in \mathbb{R}^2\;|\; p_+^2 + p_-^2 + e^{-4 \Omega}\,(V - 1) \geq 0,
\;\forall\; (p_-, p_+)\in \mathbb{R}^2 \right\}.
\end{equation}

The eigenproblem for the $\widehat{H^2}$ is equivalent to solving
the Schr{\"{o}}dinger equation for a particle in two dimensional
potential well. The difficulty in solving this equation is due to
complicated form of the potential, which would require application
of sophisticated numerical techniques. However, near the
singularity (when $\Omega \rightarrow \infty $), the potential
$e^{-4 \Omega } \left(V(\beta_\pm) - 1\right)$ can be approximated
by the hard walls equilateral triangle potential \cite{cwm2,RS}.
This can be deduced from the structure of the equipotential lines
of $V$ (see Fig.  \ref{figure1}) as well as from the asymptotic
form of the potential $V \sim \frac{1}{3}e^{-8\beta_+}$ for
$\beta_{+}\rightarrow -\infty$. The effective triangle in which
the motion takes place is found to be \cite{cwm2}: $\triangle :=
\left\{ \beta_+ \geq -\frac{1}{2}\Omega, \beta_{-}\leq
-\frac{\sqrt{3}}{3}(\beta_{+}-\Omega), \beta_{-}\geq
\frac{\sqrt{3}}{3}(\beta_{+}-\Omega)\right\} $.

Employing the hard wall approximation, the operator
$\widehat{H^2}$ given by (\ref{ap6}) reduces to
\begin{equation}\label{Htriangle}
\widehat{H_{\triangle}^2} = -\hslash^2\Big(\frac{\partial^2}{\partial \beta^2_+}
+\frac{\partial^2}{\partial \beta^2_-}\Big),
\end{equation}
with the domain
\begin{equation}
D(\widehat{H_{\triangle}^2}):= \left\{ \phi \in \mathcal{H}_{\triangle}, \left.
\phi \right|_{\partial \triangle}
=0 \right\}. \label{HtriangDom}
\end{equation}
The Hilbert space $\mathcal{H}_{\triangle}= L^2\left( \triangle,
d\beta_{+}d\beta_{-}\right)$ has the inner product $ \langle \phi
| \psi \rangle = \int_{\triangle} \bar{\phi}\psi
d\beta_{+}d\beta_{-}$. One can show that the operator
(\ref{Htriangle}) defined on (\ref{HtriangDom}) is self-adjoint. 
Thus, the eigenvalues of the operator
$\widehat{H_{\triangle}^2}$ are real and the corresponding
eigenvectors form an orthogonal set. Despite of non-separability,
the eigenequation $\widehat{H_{\triangle}^2}| q,p \rangle
=e^2_{q,p} | q,p \rangle$  can be solved analytically, leading to
the following set of eigenvalues \cite{WKL}:
\begin{equation}
e^2_{q,p} = (p^2+pq+q^2)E_0,
\label{H2spect}
\end{equation}
where $E_0 = \frac{4}{9} \frac{h^2}{\Omega^2} $, and the eigenvalues are labeled by
\begin{equation}
q = \left\{  \begin{array}{c}  0,1,2, \dots,  \\ 1,2,3, \dots,
\\  \frac{1}{3},\frac{2}{3},\frac{4}{3},\frac{5}{3}, \dots,
\end{array}   \right.
\end{equation}
and where
\begin{equation}
p = q+1, q+2, \dots.
\end{equation}

In what follows we are interested in the spectrum of the operator
$\widehat{H_{\triangle}}$ rather than the
$\widehat{H_{\triangle}^2}$. Because of self-adjointness of the
operator $\widehat{H_{\triangle}^2}$  the spectrum of
$\widehat{H_{\triangle}}$  can be obtained from the  spectral
decomposition $\sqrt{\widehat{H^2_{\triangle}}}= \int
\sqrt{\lambda} d\widehat{P}_{\lambda}$. The operator
$\widehat{H_{\triangle}^2}$ is positive $\left(\langle
\widehat{H_{\triangle}^2}\phi| \phi \rangle \geq 0 \ \ \forall
\phi \in D(\widehat{H_{\triangle}^2})\right)$. Thus, the spectrum of
$\widehat{H_{\triangle}}$ is real and given by (up to the overall
sign):
\begin{equation}
e_{q,p} =  \sqrt{E_0} \sqrt{p^2+pq+q^2}.
\label{SpectrumSquare}
\end{equation}

\section{Chaotic dynamics}

The classical dynamics of the Bianchi IX model exhibits chaotic
behavior (see, \emph{e.g.} \cite{NJC}), reflecting
non-integrability of the system. The chaotic nature of the
classical dynamics can be observed also at the quantum level. In
particular, it has been shown that wave packets evolving in the
Bianchi IX potential undergo defragmentation characterizing
chaotic behavior \cite{TF, Berger:1989jm}.

An alternative way of looking at chaoticity of the quantum systems
is by analyzing spectra of the Hamiltonian operators. Namely, it
has been conjectured that chaotic nature of the classical dynamics
leads to the level repulsion in the quantum theory
\cite{Bohigas:1983er, FH}. This behavior can be characterized by
different statistics of the level spacing $s$ variable, with the
mean value normalized to unity.

In case of the non-integrable systems,  the level spacing
distribution is often characterized by the distribution obtained
for Gaussian Orthogonal Ensemble (GOE) of the random matrices:
\begin{eqnarray}
P_{GOE}(s) = \frac{\pi}{2} s \exp\left(- \frac{\pi}{4} s^2\right).
\label{GOE}
\end{eqnarray}
In contrast, integrable systems are characterized roughly by the
Poisson distribution
\begin{equation}
P_{Poisson}(s) = e^{-s}.
\label{Poisson}
\end{equation}

In real system, chaotic and regular regimes often coexist: the
level spacing distribution can be modeled by distribution
interpolating between Poisson and GOE distributions, such as the
Brody distribution
\begin{eqnarray}
P_{Brody}(s,\beta) =  (\beta+1)b s^{\beta} \exp\left(- b s^{\beta+1}\right),
\end{eqnarray}
where $b = \left[\Gamma\left( \frac{\beta+2}{\beta+1} \right) \right]^{\beta+1}$. 
At the limits of the interval $\beta \in[0,1]$ we have $P_{Brody}(s,0)=P_{Poisson}(s)$ 
and $P_{Brody}(s,1)=P_{GOE}(s)$.

\section{Level spacing distribution}

\begin{figure}[htb]
\centering $\begin{array}{cc}  \ \ \ \ \ \ &\ \ \ \ \  \\
a)\includegraphics[width=5.5cm, angle = 0]{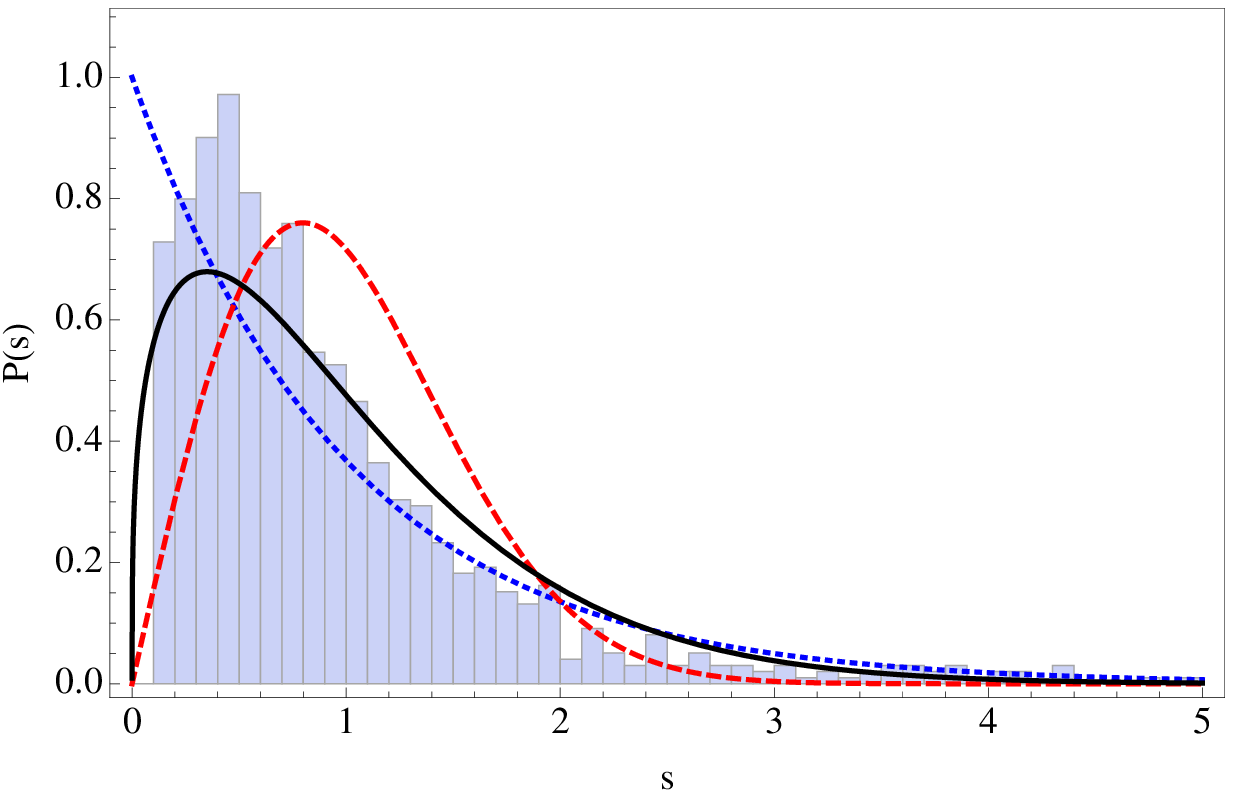}   &
b)\includegraphics[width=5.5cm, angle = 0]{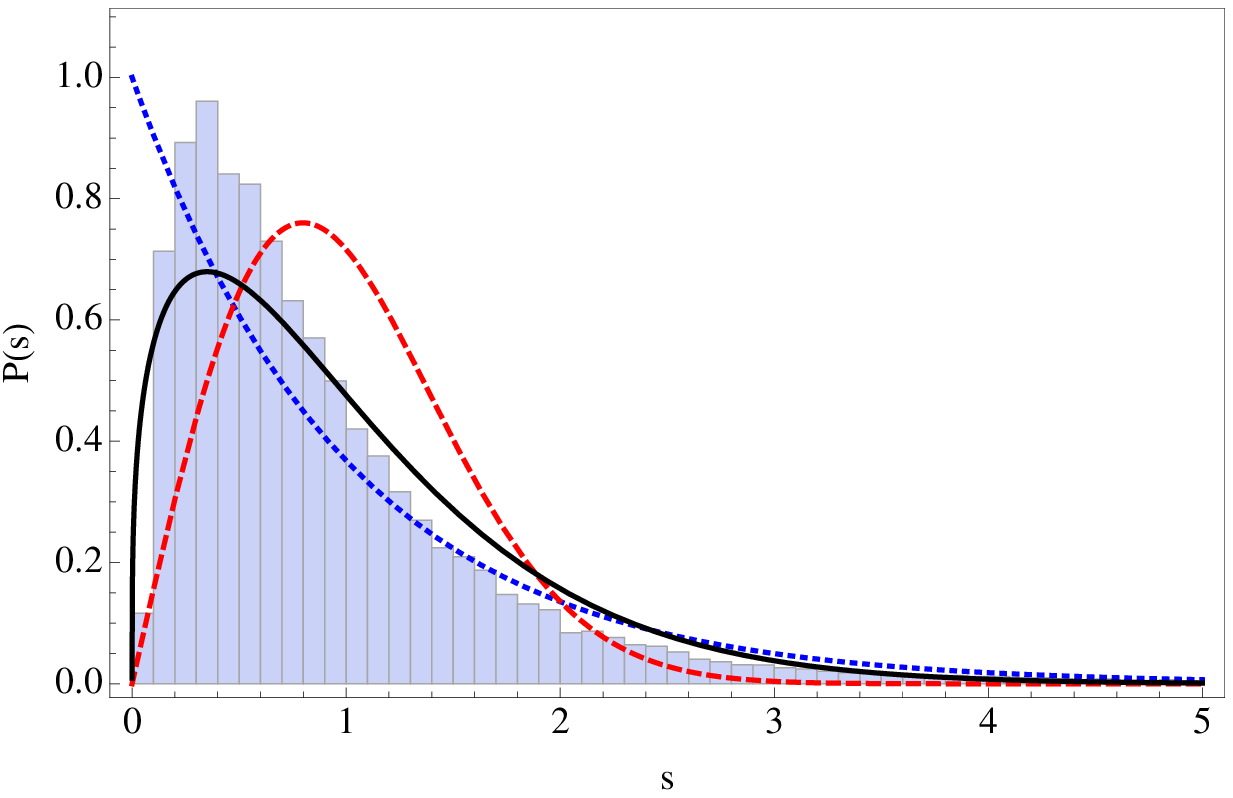} \\
\end{array}$
\caption{Level spacing distributions for the triangular hard wall
potential approximation to the Bianchi IX potential with a
different number of levels taken into account: a)  1000, b) 20000.
The dotted (blue) line corresponds to the Poisson distribution.
The dashed (red) line is the GOE distribution. The solid (black)
line is the Brody distribution with the parameter $\beta=0.3$.}
\label{Dist1}
\end{figure}

In this section, statistical properties of eigenvalues of the
prototype of the Bianchi IX with the potential approximated by the
hard wall equilateral triangle are analyzed. For this purpose
degeneracy of the spectrum (\ref{SpectrumSquare}) has to be
removed firstly. The different eigenvalues are subsequently sorted
ascending ($e_1<e_2<e_3<\dots< e_n$) and the level-spacings
$\Delta_i =e_{i+1}-e_i$ are  computed. Distribution of $\Delta_is$
cannot be compared to the scale-free distributions (\ref{GOE}) and
(\ref{Poisson}). This requires introducing normalized
level-spacing $s$, defined such that the mean value $\langle s
\rangle=1$. The simplest way  is to rescale $\Delta_i \rightarrow
s_i=  \Delta_i/\bar{\Delta}$ where $\bar{\Delta} = \frac{1}{n-1}
\sum_{i=1}^{n-1} \Delta_i$. In Fig. \ref{Dist1} we present
corresponding distributions for two different values of the number
of eigenvalues $n=1000$ and $n=20000$.

\begin{figure}[htb]
\centerline{
\includegraphics[width=8cm]{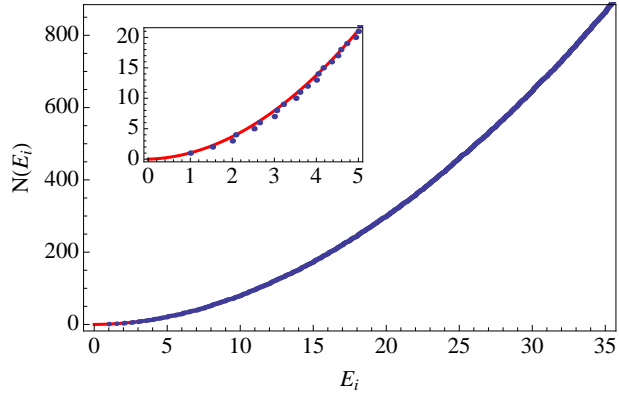}}
\caption{Plot of the function $i=N(E_i)$ together with the polynomial fit (\ref{fit}).}
\label{NE}
\end{figure}

\begin{figure}[htb]
\centering $\begin{array}{cc}  \ \ \ \ \ \ &\ \ \ \ \  \\
a)\includegraphics[width=5.5cm, angle = 0]{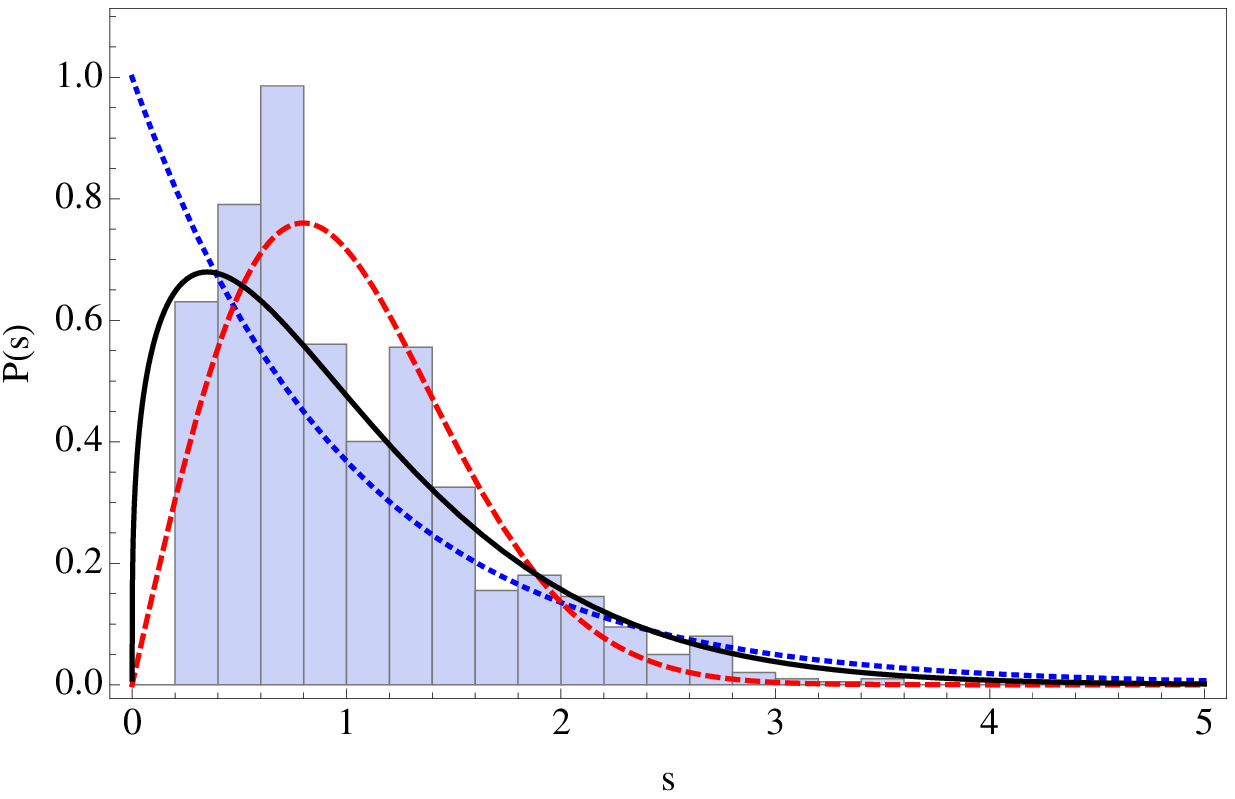}   &
b)\includegraphics[width=5.5cm, angle = 0]{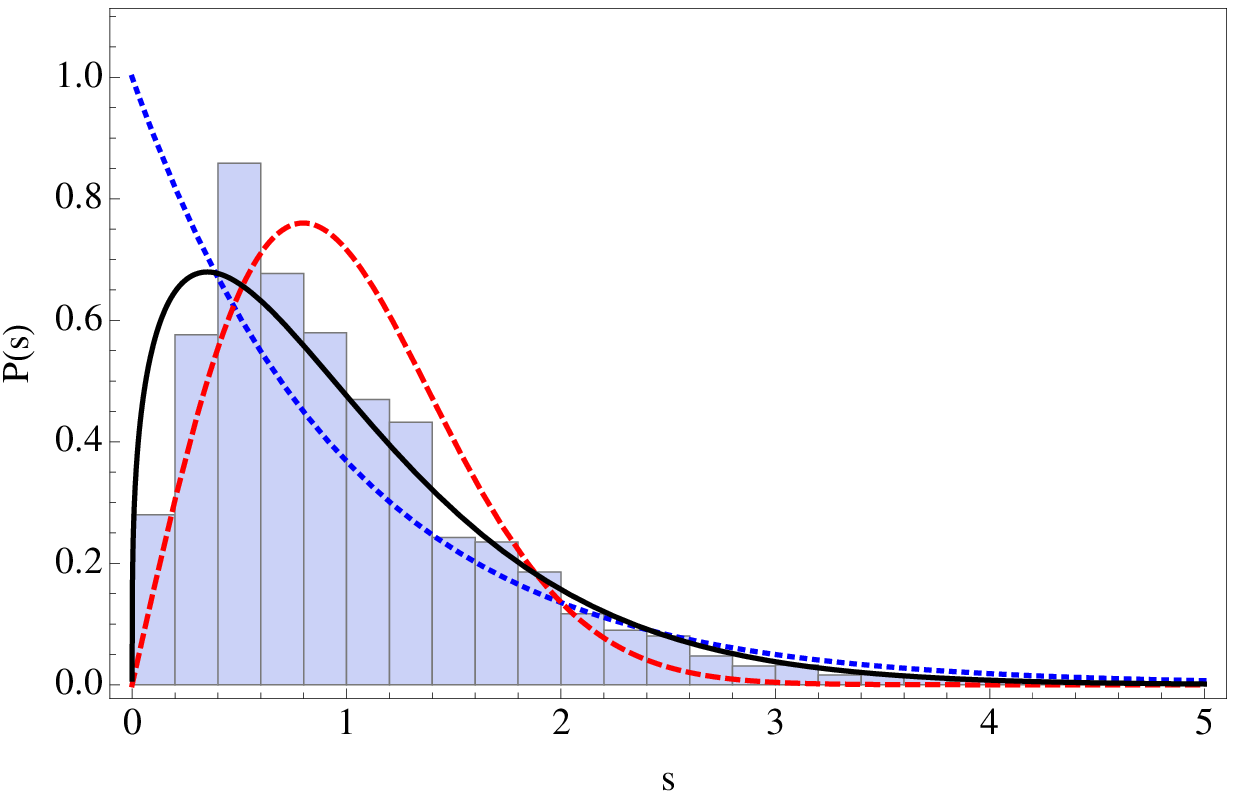} \\
\end{array}$
\caption{Level spacing distributions of the unfolded spectra for the triangular
hard wall potential approximation of the Bianchi IX potential with a
different number of levels taken into account: a)  1000, b) 20000.
The dotted (blue) line corresponds to the Poisson distribution. The dashed (red)
line is the GOE distribution. The solid (black) line is the Brody distribution
with the parameter $\beta=0.3$.}
\label{Dist2}
\end{figure}

The spectrum of eigenvalues is usually inhomogeneous (averages over
its different regions do not overlap). In order to extract information about
the local correlations, global trend in the spectrum has to be subtracted.
This can be performed by applying the so-called \emph{unfolding} of the
spectrum. In what follows we apply the procedure to the spectrum (\ref{SpectrumSquare}).
For this purpose we define $E_i \equiv E_{q,p} := \sqrt{p^2+pq+q^2}$ and (as previously)
construct non-degenerate spectrum from it. Based on this, the function $N(x)$
is defined such that $i=N(E_i)$. The function together with the polynomial fit
\begin{equation}
\bar{N}(x) = 0.209129 x + 0.830051 x^2 - 0.00542089 x^3 + 0.000050016 x^4,
\label{fit}
\end{equation}
is shown in Fig. \ref{NE}.  The fit has been performed using the least-squares method for
$n=1000$.
The fitted function $\bar{N}(x)$ allows us for the following
decomposition $N(E_i)= \bar{N}(E_i)+\tilde{N}(E_i)$, where
$\tilde{N}(E_i)$ quantifies fluctuations around the mean value.
Now the new spectrum can be defined as $\epsilon_i :=
\bar{N}(E_i)$. The advantage of such definition is that the
corresponding distribution of level spacing $s_i = \epsilon_{i+1}-\epsilon_{i}$ 
is well normalized, because
\begin{equation}
\langle s \rangle = \frac{s_1+\dots+s_{n-1}}{n-1} =  \frac{\epsilon_n-\epsilon_1}{n-1}
= \frac{\bar{N}(E_n)-\bar{N}(E_1)}{n-1} \approx   \frac{n-1}{n-1}  =1.
\end{equation}
Indeed, for the fit (\ref{fit}) performed for $n=1000$ we obtain  $\langle s \rangle = 1.00064$.
In Fig. \ref{Dist2} we present  distributions of the unfolded spectra for two different
values of the number of eigenvalues $n=1000$ and  $n=20000$.
In this case fluctuation are greater than previously, therefore size of the bin
has been increased to $0.2$ in order to smoothen the spectrum. Furthermore,
for $n=20000$ the unfolding procedure has been performed based on the
new fit  $\bar{N}(x)=-4.82423 x + 1.23144 x^2 - 0.016761 x^3 + 0.00021306 x^4 -
 1.36681\cdot10^-6 x^5 + 4.09267\cdot10^{-9} x^6 - 4.56953\cdot10^{-12} x^7$,
which gives $\langle s \rangle = 1.00079$. The fit (\ref{fit}) performed for $n=1000$
is no more valid for the $n=20000$ case.

\section{Summary}

The level spacing distributions obtained with and without applying the unfolding
procedure are qualitatively similar. This indicates that inhomogeneity of the spectrum
does not play a significant role. Both spectra are qualitatively described by the
Brody's distribution with the parameter $\beta \approx 0.3$. The observed deficit
of the small level spacing (with respect to the Poisson distribution) indicates for
level repulsion, which might reflect chaotic nature of Bianchi IX model. Confirmation
of this observation might be provided by the further analysis of fluctuations of the
unfolded spectra \cite{Relano}. Furthermore, next step would be numerical
examination of the level statistics of the Bianchi IX with the exact potential.

\vspace{1.0cm}
%\begin{acknowledgments}
\noindent {\it Acknowledgments.} W.P. would like to thank the
organizers of the Conference for creating an inspiring atmosphere,
and for financial support.
%\end{acknowledgments}


\begin{thebibliography}{99}

\bibitem{LK} E.~M.~Lifshitz and I.~M.~Khalatnikov,  Adv.\ Phys.\ {\bf 12} (1963) 185.

\bibitem{BKL1}
V.~A.~Belinsky, I.~M.~Khalatnikov, E.~M.~Lifshitz,
Adv.\ Phys.\  {\bf 19} (1970) 525.

\bibitem{BKL2}
 V.~A.~Belinsky, I.~M.~Khalatnikov, E.~M.~Lifshitz, Adv.\ Phys.\  {\bf 31} (1982) 639.

\bibitem{cwm2} C. W. Misner, ``Quantum Cosmology'',
Phys.\  Rev.\  {\bf 186},  1319 (1969).

\bibitem{NJC} N.~J.~Cornish and J.~J.~Levin,
Phys.\ Rev.\ Lett.\  {\bf 78} (1997) 998; Phys.\ Rev.\ D {\bf 55}
(1997) 7489

\bibitem{DHK} D.~H.~King, Phys.\ Rev.\ D {\bf 44}, 2356 (1991).

\bibitem{cwm3} C. W. Misner, ``Minisuperspace'', in {\it Magic Without
Magic: John Archibald Wheeler}, p. 441, Ed. by J. R. Klauder (W.H.
Freeman and Company, San Francisco, 1972).


\bibitem{RS} M. P. Ryan Jr., and L. C. Sheply, {\it Homogeneous Relativistic
Cosmologies} (Princeton University Press, Princeton, 1975).

\bibitem{WKL} Wai-Kee Li, S.M. Blinder,
%``Solution of the Schr\"{o }dingerequation for a particle in a equilateral triangle'',
J. Math. Phys. {\bf 26}, 2784 (1985).

\bibitem{TF} T.~Furusawa,
%``Quantum Chaos of Mixmaster Universe'',
Prog. Theor. Phys. {\bf 75} (1986) 59; {\bf 76} (1986) 67.

\bibitem{Berger:1989jm}
  B.~K.~Berger,
  %``Quantum Chaos in the Mixmaster Universe,''
  Phys.\ Rev.\ D {\bf 39} (1989) 2426.
  %%CITATION = PHRVA,D39,2426;%%
  %20 citations counted in INSPIRE as of 02 Nov 2014

%\cite{Bohigas:1983er}
\bibitem{Bohigas:1983er}
  O.~Bohigas, M.~J.~Giannoni and C.~Schmit,
  %``Characterization of chaotic quantum spectra and universality of level fluctuation laws,''
  Phys.\ Rev.\ Lett.\  {\bf 52} (1984) 1.
  %%CITATION = PRLTA,52,1;%%
  %308 citations counted in INSPIRE as of 02 Nov 2014

\bibitem{FH} F.~Haake, {\it Quantum Signatures of Chaos} (Springer, Berlin, 2001).

\bibitem{Relano} A.~Relano, J.~M.~G.~Gomez, R.~A.~Molina, J.~Retamosa, and E.~Faleiro,
Phys.\ Rev.\ Lett. {\bf 89}, 244102 (2002).

\end{thebibliography}
\end{document}